\documentclass[10pt]{article}
\usepackage{thumbpdf}
\usepackage[pdftex]{hyperref}
\usepackage{color}

\usepackage{amsmath}
\usepackage{amssymb}
\usepackage{graphicx}
\usepackage{setspace}
\usepackage[utf8]{inputenc} 
\usepackage[T1]{fontenc}
\usepackage{eepic}
\usepackage{cite}
\usepackage{url}
\usepackage{epstopdf}
\usepackage{hyperref}
\usepackage[labelfont=bf,labelsep=period,justification=raggedright]{caption}

\usepackage{color}

\makeatletter
\renewcommand{\@biblabel}[1]{\quad#1.}

\renewcommand{\subsubsection}[1]{\vskip 1em \noindent\textit{#1}.}

\newcommand{\beq}{\begin{equation}}
\newcommand{\eeq}{\end{equation}}
\newcommand{\bea}{\begin{align}}
\newcommand{\eea}{\end{align}}
\newcommand{\bN}{\bar{N}}

\makeatother

\date{}

\begin{document}

\begin{flushleft}
{\Large
\textbf{Extraction of Temporal Networks from Term Co-occurrences
in Online Textual Sources}}


Marko Popovi\'c$^{1}$, 
Hrvoje \v{S}tefan\v{c}i\'c$^{1,2}$, 
Borut Sluban$^{3}$,
Petra Kralj Novak$^{3}$,
Miha Gr\v{c}ar$^{3}$,
Igor Mozeti\v{c}$^{3}$,
Michelangelo Puliga$^{4}$,
Vinko Zlati\'c$^{1,\ast}$
\\
\bf{1}  Theoretical Physics Division, Rudjer Bo\v{s}kovi\'c Institute, P.O.Box 180, HR-10002 Zagreb, Croatia
\\
\bf{2} Catholic University of Croatia, Zagreb, Croatia
\\
\bf{3} Jo\v{z}ef Stefan Institute, Ljubljana, Slovenia
\\
\bf{4} IMT Alti Studi Lucca, Lucca, Italia
\\
$\ast$ E-mail: vinko.zlatic@irb.hr
\end{flushleft}

\section*{Abstract}
A stream of unstructured news can be a valuable source of hidden
relations between different entities, such as financial institutions, countries, or persons.
We present an approach to
continuously collect online news, recognize relevant entities in them,
and extract time-varying networks.  The nodes of the network are the
entities, and the links are their co-occurrences.  We present a method
to estimate the significance of co-occurrences, and a benchmark model
against which their robustness 
is evaluated.
The approach is applied to a large set of financial news, collected over a period of two years.
The entities we consider are 50 countries which issue sovereign bonds,
and which are insured by Credit Default Swaps (CDS) in turn.
We compare the country co-occurrence networks to the CDS networks 
constructed from the correlations between the CDS.
The results show relatively small, but significant overlap between
the networks extracted from the news and those from the CDS correlations.


\section*{Introduction}

During the last decade, methods developed in the fields of 
mathematics, computer science and statistical physics have contributed
to the emergence of complex networks analyses. These analyses have
strongly penetrated into the 
 areas of social media, biology,
and economics\cite{caldarelli2007scale, jackson2010social}. 
A special type of networks extracted from data are co-occurrence networks,
 used in diverse fields, such as: 
linguistics\cite{edmonds1997choosing}, 
bioinformatics\cite{cohen2005using,wilkinson2004method,shalgi2007global},
ecology\cite{freilich2010large},
scientometry\cite{su2010mapping,mane2004mapping}, and socio-technological 
networks\cite{cattuto2007network, zlatic2009hypergraph, ghoshal2009random}. 
Co-occurrence networks are
loosely defined as networks in which nodes represent some entities
(for example persons, companies, genes, etc.),
and links represent the fact that
these entities exist together in some collection (for example
database, article, etc.). For textual sources it is of paramount
importance to extract the links between entities that 
represent a real relationship and 
are not created by chance.
Furthermore, beside reliability,
it is important  that the extraction of reliable co-occurrences
is implemented by an efficient algorithm. In the
case of online textual sources, the stream of data 
can be potentially large and fast, and the speed of processing
can be a decisive factor in  the choice between
alternative methods. 

This paper addresses the question of reliable and efficient
construction of co-occurrence networks from textual sources on the web.
The main result is a significance algorithm, based on a simple 
algebraic method and counting statistics, that can be efficiently 
used to extract significant co-occurrences in the real data stream.
Another result is a benchmark model, used to generate 
synthetic data, on which the significance algorithm is tested and
the required parameters are determined. 
Finally, we investigate the relation between the networks extracted 
from online texts and the networks drawn from economic data.
We demonstrate an application of our method by extracting a network of co-occurring countries from financial news.

Alternative methods to construct co-occurrence networks 
have previously been used\cite{caldarelli2007scale}.
The most common is the so called
Maslov-Sneppen rewiring algorithm\cite{maslov2002specificity} which is
known to produce randomized networks and is a ``microcanonical''
alternative to the ``canonical'' configuration models, such as the Molloy Reed
model\cite{molloy1995critical}. These methods create randomized
versions of initial networks in which degrees are conserved and
afterward a comparison with real data is made to check if certain
patterns differ significantly from those obtained by the randomization
procedure. These methods can be cumbersome, especially if the frequency
of temporal changes in the network is very high.

Benchamrk model which is presented in the paper is a simple linear hidden variable model that we use to construct syntethic data. These data are not a representative of the real co-occurence data and their statistics is not the one we measure in co-occurence analysis. This data just have the same structure (i.e. projection of bipartite graph) and through hidden variables we can construct relationships as important or nonimportant in order to evaluate how well our method performs with respect to different measurement parameters.

The case study, which is an integral part of the paper, 
 analyzes textual data collected from 2,500 RSS feeds from 170  major 
English-language news web sites, with the subject of economy and finance. We acquired around 35,000 articles per day, during a period of two years, about 18 million articles in total. The data acquisition
pipeline  processes the data in real-time, and can easily be
extended to other news sources and languages. In that case, the number of processed articles could increase
many fold. Therefore, it is vital to have a method which can extract significant co-occurrences from a large dataset efficiently.
In the paper we present an efficient algebraic method that can extract co-occurrences from simple counting statistics.
We believe that the method can be incorporated into the real-time
acquisition pipeline, but we do not address the issue of efficiency in the paper.

Other types of network analyses of online textual sources such as
Twitter, Facebook, Google, etc. are also an interesting topic of
research. In particular, the idea that one can extract some unaccounted economical/financial information related to some entity from the
online text materials, even before a market can account for it, has
been investigated thoroughly \cite{bollen2011twitter, mishne2006predicting, de2008can, asur2010predicting, ruiz2012correlating}. Several economic indicators can also be represented as network measures, such as the impacting-impacted vulnerability derived from  CDS networks of companies \cite{kaushik2013credit}. Therefore in the end, we use our method to find significant co-occurences of countries in the news web sites. We then consider CDS (Credit Default Swaps) of the same countries, and extract networks based on the correlations between CDS time series over the same time windows.
The results show relatively small, but significant overlap between the networks extracted from the news and those from CDS correlations.


This paper is organized as follows. First, we describe the textual
data that we use and the architecture of the real-time data acquisition pipeline.
We then describe the method
for extraction of significant co-occurrences. 
Further, we develop a benchmark model for 
the creation of realistic synthetic data. 
We test our extraction algorithm on the benchmark model in order to estimate the
statistics we need for the method to work reliably. In the case study, we
construct co-occurrence networks from the textual data acquired, 
and compare them with the CDS networks. 
Finally, we give conclusions and state what further developments we might achieve with the
methods described.

\section*{Data Acquisition}
\label{sec:dacq}
This section briefly describes the technology needed to extract
bipartite networks from textual sources on the web.
The idea is to monitor a large number of financial data sources
(news and blog sites), acquire their content, extract relevant
entities, and construct networks in different time windows.
Within a specific time window (e.g., a month), nodes of the
network are all the entities of interest (e.g., financial
institutions or countries) which appear in the
texts, and links are formed by their co-occurrences in the
same documents. 

\begin{figure}[!ht]
\begin{center}
\includegraphics[width=6in]{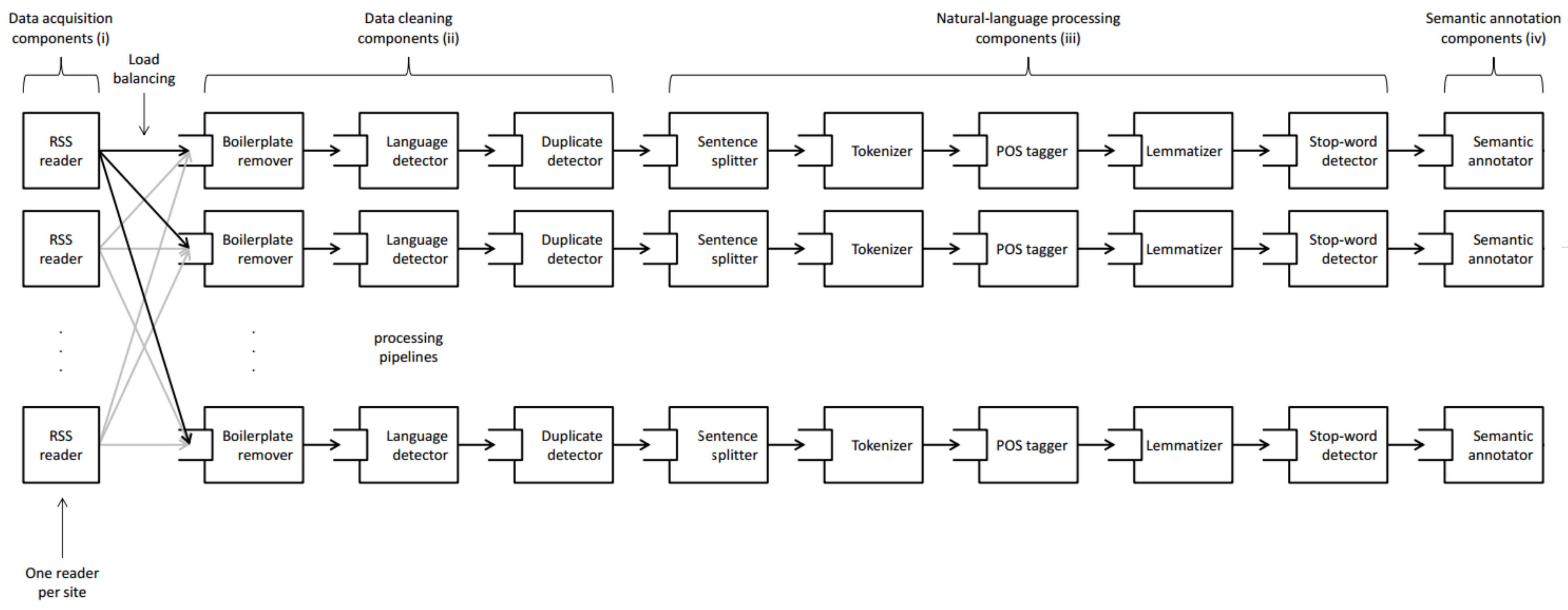}
\end{center}
\caption{{\bf The data acquisition, processing and semantic annotation pipeline.}}
\label{fig:pipeline}
\end{figure}

The technology required for network construction is implemented as a
data acquisition and processing pipeline (DacqPipe in short).  It is
responsible for acquiring unstructured data from several data sources,
preparing it for the analysis, and brokering it to the appropriate
analytical components. The DacqPipe is running continuously, since
October 2011, polling the web and proprietary APIs for recent content,
turning it into a stream of preprocessed text documents. It is
composed of two main parts: the data acquisition and the semantic data
processing. The pipeline is schematically presented in
Figure~\ref{fig:pipeline}.

\subsection*{The data acquisition pipeline}
The news articles and blogs are collected from 2,503 RSS feeds from 170 English language web sites (14,567 domains), covering the majority of web news in
English and focusing on financial news and blog sources. We collect
data from the main news providers and aggregators (like yahoo.com,
dailymail.co.uk, nytimes.com, bbc.co.uk, wsj.com) and also from the
main financial blogs (like zerohedge.com). The fifty most productive
web sites account for 80\% of the collected documents.

We started with continuous data acquisition on October 24, 2011. In the period from
November 2011 until the end of 2013, almost 18 million documents were
collected and processed.  On an average work day, about 40,000 unique 
articles are collected. The number of collected articles is
substantially lower during weekends; around 20,000 per weekend
day. Holidays are also characterized by a lower number of documents.

Content from news, blogs, forums, and other web content, is not
immediately ready to be processed by the text analysis methods. Web
pages contain a lot of `noise' or `boilerplate' (i.e., undesired
content such as advertisements, copyright notices, navigation
elements, and recommendations) that needs to be identified and removed
before the content can be analyzed. For this reason, the data
acquisition and preprocessing pipeline (DacqPipe) consists of a number
of components: (i) data
acquisition components, (ii) data cleaning components, (iii)
natural-language preprocessing components and (iv) semantic annotation
components.  The pipeline topology is shown in
Figure~\ref{fig:pipeline}.

The data acquisition components are mainly RSS readers that poll for
data in parallel. One RSS reader is instantiated for each web site of
interest. The RSS sources, corresponding to a particular web site, are
polled one after another by the same RSS reader to prevent the servers
from rejecting requests due to concurrency. An RSS reader, after it
has collected a new set of documents from an RSS source, dispatches
the data to one of several processing pipelines. The pipeline is
chosen according to its current load (load balancing). A processing
pipeline consists of a boilerplate remover, duplicate detector,
language detector, sentence splitter, tokenizer, part-of-speech
tagger, lemmatizer, stop-word detector and a semantic annotator. Some
of the components are custom-made while others use the functionality
available from the OpenNLP library~\cite{OpenNLPWebsite}.

\subsubsection{Boilerplate Remover}
Extracting meaningful content from web pages presents a challenging
problem which was extensively addressed in the static setting. Our
setting, however, is dynamic and focuses on content extraction from streams of HTML
documents in real time. 
We use the URL Tree content extraction algorithm \cite{SlubanBoilerplate}, which 
is specialized for content extraction from streams of HTML documents.
The algorithm is based on the observation that HTML documents 
from the same source normally share a common template. 
The content extraction algorithm is efficient, unsupervised, and language-independent.

\subsubsection{Duplicate Detector}
Due to news aggregators and redirect URLs, one article can appear on
the web with many different URLs pointing to it. To have a concise
dataset of unique articles, we have developed a duplicate detector
that is able to detect if the document was already acquired or not.

\subsubsection{Language Detector}
It detects the language used in a document and discards
all the non-English documents. The model is constructed by a machine learning
algorithm, and trained on a large multilingual set of documents. The basic features
for model training are the frequencies of several consecutive letters.

\subsubsection{Sentence Splitter}
The sentence splitter splits the text into sentences. The result forms the input to the part-of-speech tagger. We use the
OpenNLP~\cite{OpenNLPWebsite} implementation of the sentence splitter.

\subsubsection{Tokenizer}
Tokenization is the process of breaking a stream of text into words,
phrases, symbols, or other meaningful elements called tokens. In DacqPipe
our own implementation of the tokenizer is used, which
supports the Unicode character set and is based on rules.

\subsubsection{Part-of-Speech Tagger}
The part-of-speech (POS) tagger marks tokens with their corresponding
word type (e.g., noun, verb, proposition) based on the token itself
and the context of the token. A token might have multiple POS tags
depending on the token and the context. The part-of-speech tagger from
the OpenNLP library~\cite{OpenNLPWebsite} is used.

\subsubsection{Lemmatizer}
Lemmatization is the process of finding the normalized forms of words
appearing in text. It is a useful preprocessing step for a number of
language engineering and text mining tasks, and especially important
for languages with rich inflectional morphology. In DacqPipe, we use
LemmaGen~\cite{JursicEtAl-LemaGen2010} for lemmatization, which is the
most efficient publicly available lemmatizer trained on large lexicons
of multiple languages, whose learning engine can be retrained to
effectively generate lemmatizers of other languages. We lemmatize to
English.

\subsubsection{Stop-word Detector}
In automated text processing, stop words are words that do not carry
semantic meaning. In DacqPipe, stop words are
detected and annotated.

\subsection*{Semantic Data Processing}
The data acquisition pipeline is general, domain independent, and
biased towards finance only by the selection of RSS sources.  On the
other hand, the semantic data processing pipeline is tailored to
finance by an lightweight ontology of financial entities and terms.
The ontology includes a dictionary of positive and negative words
for dictionary-based sentiment analysis.  The ontology contains
gazetteers, which specify the lexicographic information about the
possible appearances of entities in text.  This information is used by
the semantic annotator to annotate the entities in text.

\subsubsection{Ontology of Financial Entities and Terms}
The lightweight ontology of financial entities and concepts
consists of three main categories: financial entities, financial
terms, related to the latest financial crisis, and geographical
entities. Most of the information extraction ontology is automatically
induced by reusing various data sources.  The geographical entities
(continents, countries, cities, organizations (such as European Union
and United Nations) and currencies), and the relations between those
were extracted from GeoNames\footnote{GeoNames:
\url{http://www.geonames.org/}}.  We used the IDMS database and MSN
Money\footnote{MSN Money: \url{http://money.msn.com/}} to `grow' the
ontology from a list of seed stock indices to its constituents
(stocks) and further on to the companies that issue these stocks.  We
added a list of `over-the-counter' stocks from OTC
Markets\footnote{OTC Markets: \url{http://www.otcmarkets.com/home}}.
The hierarchy of financial terms related to the financial crisis was
developed in collaboration with experts in economics.  It includes the main
European politicians and economy leaders, Central Banks and other
financial institutions, rating agencies, and fiscal and monetary policy
terms.

\subsubsection{Semantic Annotator}
Each entity has associated gazetteers; gazetteers are rules describing
the appearance of an entity in text. For example, 'The United States
of America` can appear in text as `USA', `US', `The United States' and
so on. The rules include capitalization, lemmatization, POS tag
constraints, and must contain constraints (another gazetteer must be
detected in the document or in the sentence) and followed by
constraints.

\subsubsection{Semantic Annotation Database}
The information about the location (specific paragraph) of terms
(entities and sentiment words) in each document is stored in an SQL
database. Additionally, the entity-class relationship and the
hierarchy of ontology classes is also stored. Meta-data about
the document, including the document title, acquisition and
publication time, source domain, response URL, among others, allow 
drill-down to the concrete document.  Some aggregates, like the
sentiment polarity of each document and each paragraph, are also
precomputed and stored in the database for performance reasons.  Such
a database allows for efficient and diverse querying. For example,
document titles and response URLs, documents by sources, dates, entity
content, sentiment word content and aggregate sentiment by
documents/paragraphs.

\section*{Significance Algorithm}
\label{Method}
The data thus acquired can be naturally represented as a dynamic
bipartite network. In this representation, entities and documents are
represented by two classes of nodes. Documents are used as 'dummy' nodes since we focus on
interesting relationships between entities. Documents have time stamps
which enable the extraction of entity relationships as a function of time.

When other sources of data are scarce, an entity projection network can provide important
information and present a way to infer the structure of entity
interrelationships. We devise a simple method to extract the  
relationships between entities based on their co-occurrence statistics. The method
recognizes entities as related if their co-occurrence in the data is
significantly more frequent then expected from a suitable null model, using a
level of significance as a parameter in the method. The null model is based on
entity occurrence data and is very similar to the well-known configuration model,
but is much faster to compute. This feature is important when large amounts
of data are streamed in real time, as in our case.

Data structure used in the algorithm is organized as documents with timestamps and a list of entities in each document. Time stamps allow the documents to be grouped in days, weeks or other suitable time frames. Each frame is analyzed independently. A shorter time frame provides a better temporal resolution although the frequency of events can sometimes be insufficient for reasonable significance testing. On the other hand, longer time frames will generally provide enough events for statistical testing, but a temporal resolution can be too coarse-grained for the intended purposes. The question of time frames will be further addressed in the section on benchmark models.

Some entities are more frequent in the data than others. We are interested in relations between pairs of entities and not their individual properties, therefore, we take the numbers of occurrences as an external parameter.
It must be stressed that methods and models presented in this paper can be used to create projection networks out of any temporal bipartite network. 
Two entities can appear in the same document even if they do not have any real connection. Therefore, we want to calculate, given the number of occurrences, whether two entities appear more often together than expected by chance. The standard method is to use configuration or a Maslov-Sneppen rewiring model as null-models. Here we devise a simple analytical formula based on the configuration model, to compute a significance score as a function of the number of occurrences and the total number of documents. This formula is easy to compute and is much faster than alternative Monte Carlo simulations.  In the not-unrealistic case of huge data flowing through the presented pipeline, the speed of computation becomes of paramount importance. 

In our method, considering two entities $A$ and $B$ with $N_A$ and $N_B$ occurrences respectively, we count all possible configurations in which these entities can be arranged:
\begin{align}
\binom{N}{N_A}\binom{N_A}{N_{AB}}\binom{N-N_A}{N_B-N_{AB}} 
\end{align}
where $N_{AB}$ is the number of $A$ and $B$ co-occurrences.  In the data analysis scheme we discard documents with only one entity in order to get proper statistics. 

To write down the probability of having $N_{AB}$ co-occurrences we need to normalize the above expression with sum of all possible configurations over all possible co-occurrence values. This is equal to a number of ways we can put $A$ and $B$ independently in $N$ documents:
\begin{equation}
\binom{N}{N_A}\binom{N}{N_B}
\end{equation} 

Expected number of co-occurrences is therefore given by:
\begin{align}
\langle N_{AB} \rangle &= \frac{1}{\binom{N}{N_A}\binom{N}{N_B}}\sum_{N_{AB}=0}^{\min(N_A,N_B)}N_{AB}\binom{N}{N_A}\binom{N_A}{N_{AB}}\binom{N-N_A}{N_B-N_{AB}} ,\\
&= \frac{N_A N_B}{N}
\end{align}
where $T$ is just the normalization constant, and the second moment is
\begin{align}
\langle N_{AB}^2 \rangle = \frac{N_A N_B}{N}\frac{N_A N_B + N - N_A - N_B}{N - 1}		
\end{align}
where both sums have been carried out in Mathematica.\\

Standard deviation is now 
\begin{align}
\sigma_{AB} = \sqrt{\frac{N_A N_B}{N}\left(\frac{N^2 - N(N_A+N_B) + N_A N_B}{N(N-1)}\right)}
\end{align}
and we compute a standard significance score of the co-occurrence $N_{AB}$ from the data as 
\begin{align}
Z_{AB} = \frac{N_{AB}-\langle N_{AB} \rangle}{\sigma_{AB}}
\end{align}
		
Setting some fixed threshold $Z_0$ we can distinguish significant $Z > Z_0$ and non-significant $Z < Z_0$ relationships between the entities.

\begin{figure}[!ht]
\begin{center}
\includegraphics[width=4in]{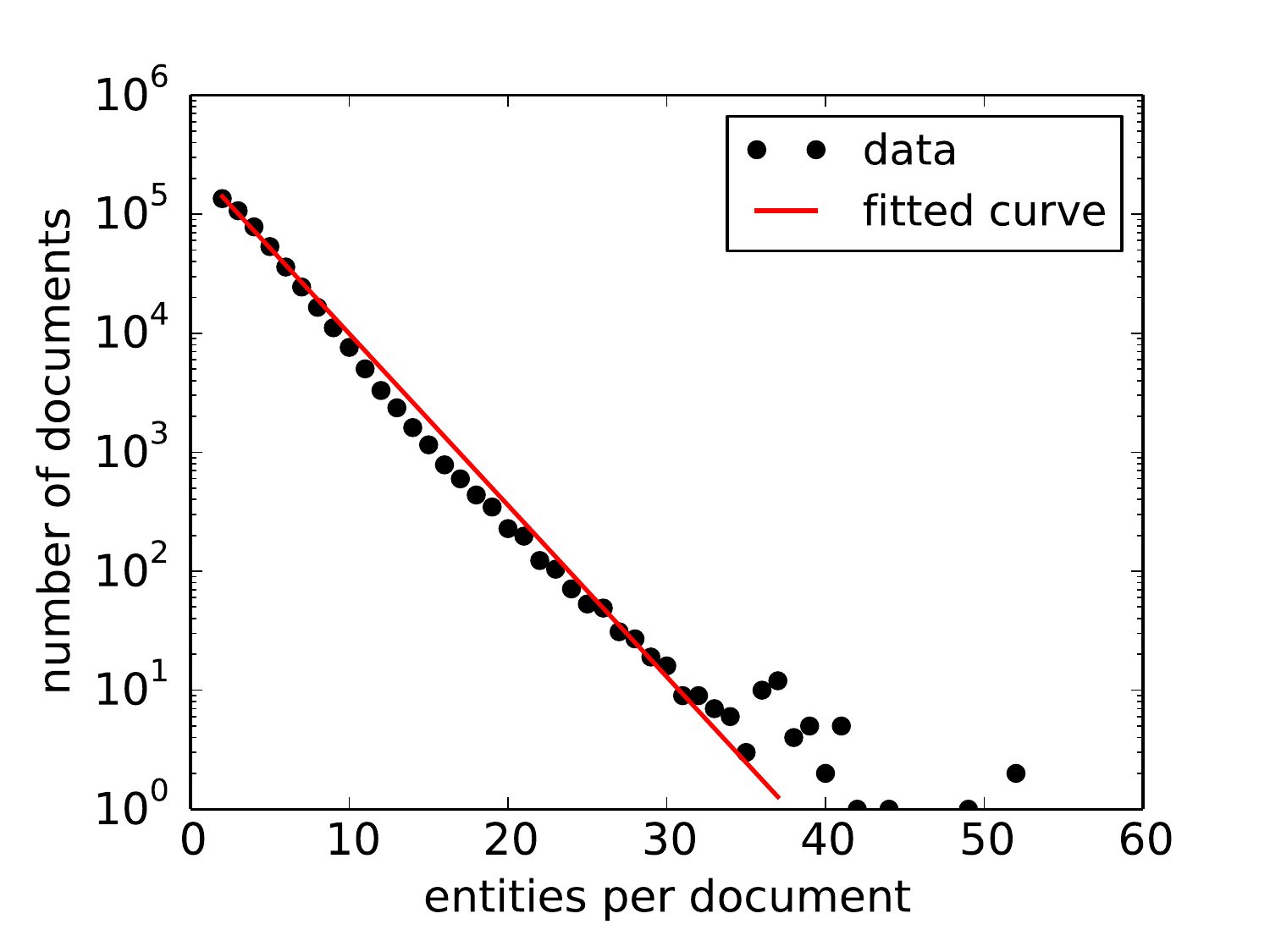}
\end{center}
\caption{{\bf Frequencies of the number of entities in the document.} The distribution has exponential tail which enforces approximations used in the paper. The exponent of the distribution is: $-0.332 \pm 0.006$ .}
\label{NumberOfEntities}
\end{figure}

\begin{figure}[!ht]
\begin{center}
\includegraphics[width=4in]{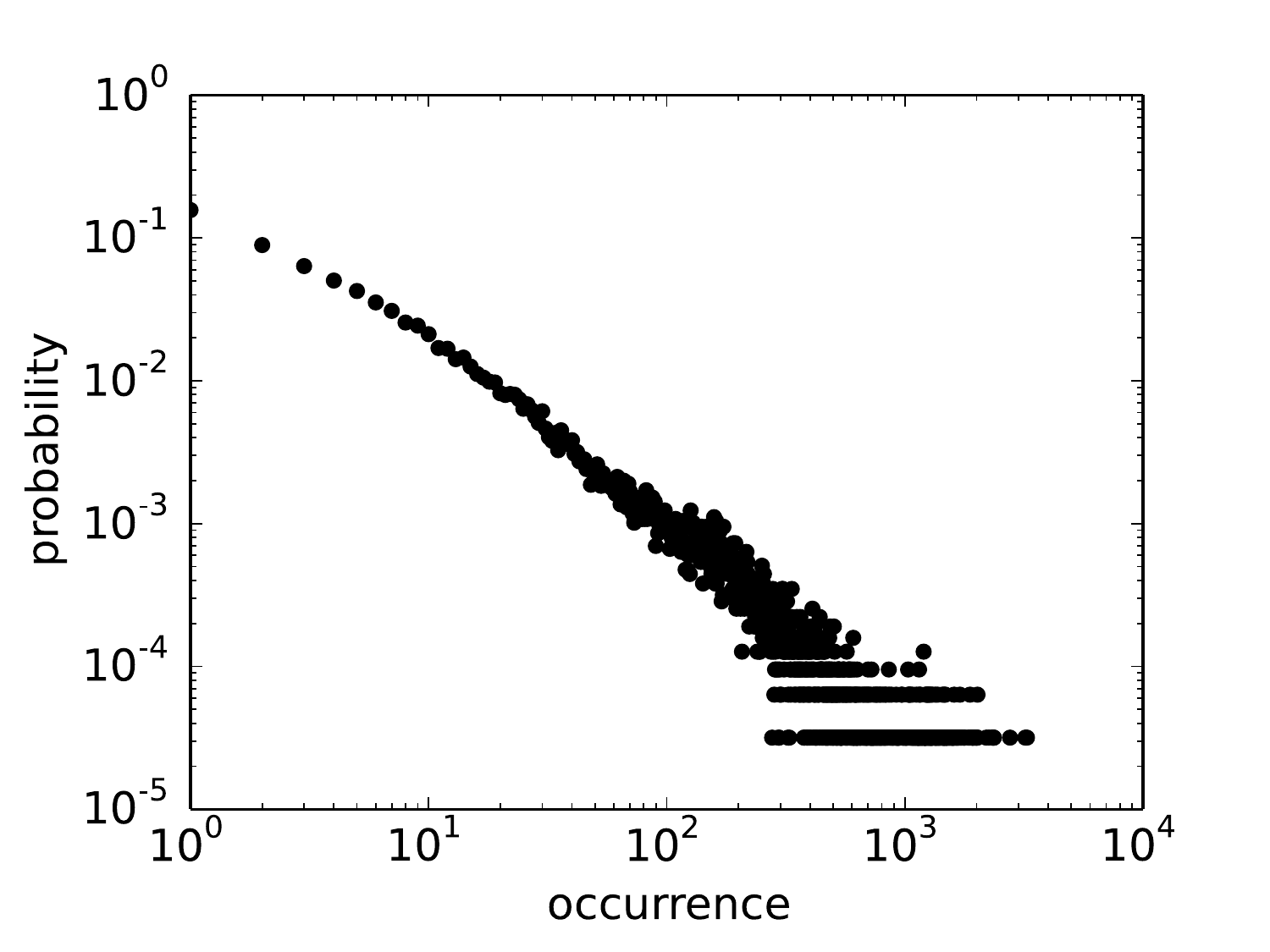}
\end{center}
\caption{{\bf Distribution of occurrence of entities in the data has fat tail. }}
\label{occurencies}
\end{figure}

\begin{figure}[!ht]
\begin{center}
\includegraphics[width=4in]{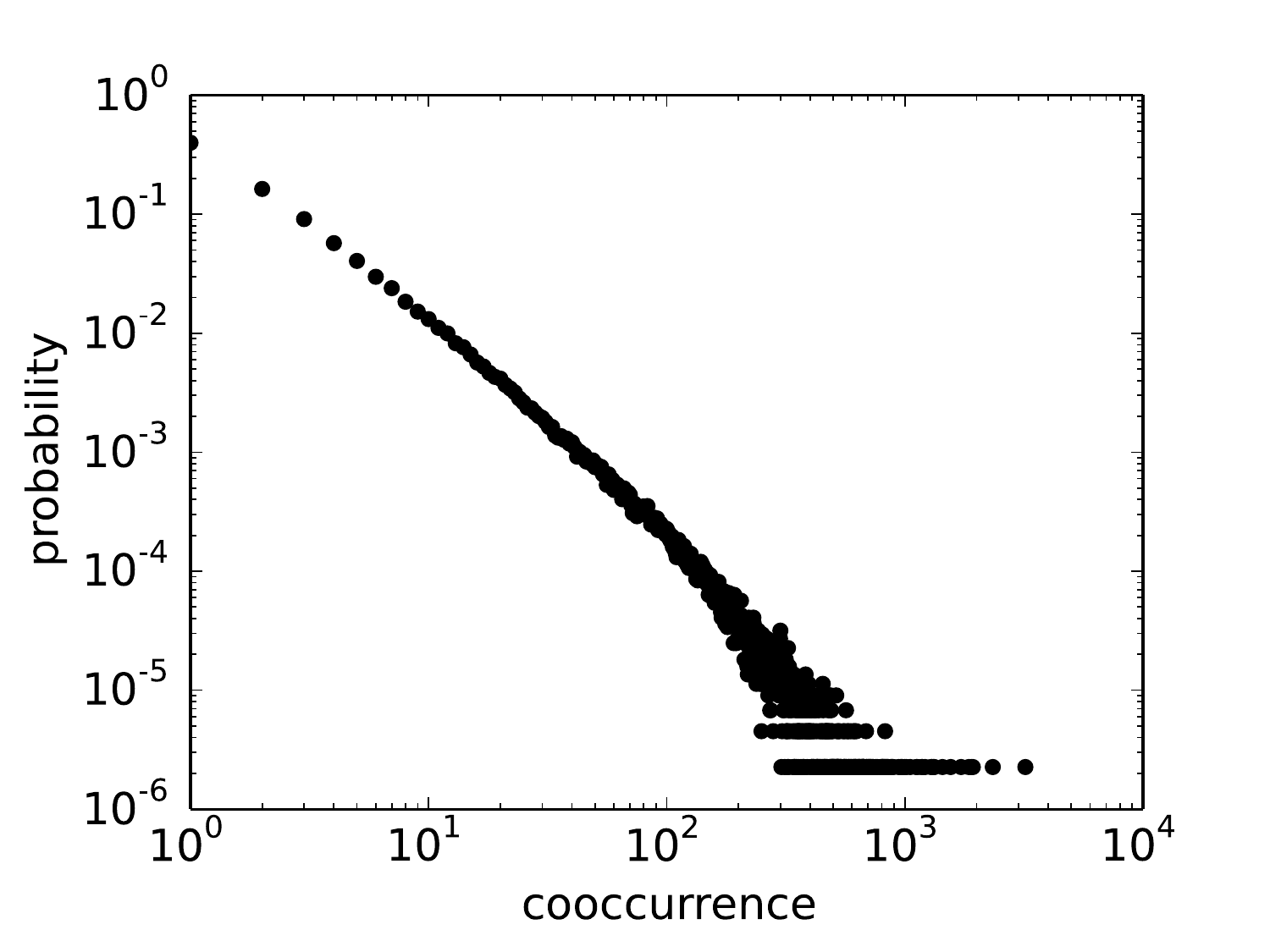}
\end{center}
\caption{{\bf Distribution of co-occurrences in the data has fat tail distribution.} }
\label{Coocurencies}
\end{figure}

This method is different from the configuration model in that we treat documents as identical 'passive' containers. As a consequence, we cannot choose the distribution of the number of entities per document since it is already determined by the algorithm. Therefore, we can expect results similar to the configuration model only if the number of entities per document in the data already has a distribution similar to the one coming out of our algorithm. We can estimate the distribution given by our algorithm in the following way. First, let us find the distribution when multiple occurrences of the same entity in a document are allowed. The number of entities per documents obeys Poisson distribution
\begin{align}
P(N_{\text{entities} = k}) &= \frac{\lambda}{k!}e^{-\lambda}
\end{align}
where $\lambda = \frac{\sum\limits_A N_A}{N}$. Now, we have to remove all configurations in which there are multiple occurrences of the same entity. This procedure prefers configurations that even more equally distribute the number of entities than predicted by a Poisson distribution. Therefore, describing all documents by average value $\frac{\sum\limits_A N_A}{N}$ should provide reasonable conditions for distributions in data.
In our data, the distribution of the number of entities per document is approximately exponential, as seen in Figure \ref{NumberOfEntities}. Since all statistical moments are non-divergent for such a distribution, the average value is a good first order representative of the data set which coincides with the algorithm condition. This method has also recently been used by \cite{tibely2013extracting}.
We also add that the individual probability distribution that an entity occurs in the document or that 2 entities co-occur in a document are heavy-tailed distributions, as seen in Figure~\ref{occurencies} and Figure~\ref{Coocurencies}.

The main advantage of this method is that it is very fast to compute. This is ideal for the huge amount of real life data and is not necessarily useful only in our context.

\section*{Benchmark model for the creation of synthetic data}

In the previous section we have defined a method that we use to find significant co-occurrences of entities in documents. If there are no underlying assumptions of the hidden relationships between these entities that are worthy of further investigation, then such an endeavor is unnecessary. For example: one can extract co-occurrences of the names of proteins from the corpus of biomedical papers hoping that these co-occurrences are related to real protein-protein interactions in the cell\cite{guldener2006mpact}. One can also extract co-occurrences of countries in financial news hoping that they will correspond to the real financial riskiness of these countries. Clearly, hidden relationships cannot be explicitly measured in the data and the level to which significant co-occurrences correspond to these relationships is impossible to estimate. To overcome such difficulties we create a simple model of hidden relationships in the spirit of the many hidden variable models \cite{GuidoFitness,popovic2012geometric}. 
In this model we \emph{explicitly} provide hidden relationships which we call \emph{importances} and use them to construct an ensemble of artificial networks. Our goal is to investigate how many of the important relationships will be discovered by the method with respect to different statistical parameters. The main idea behind this hidden variable model is the one which is heavily used in community finding \cite{lancichinetti2008benchmark,fortunato2010community}, and is used to evaluate different community finding algorithms with respect to their performance.

We therefore present a simple but broadly applicable benchmark model to test robustness
and predictability of the described method. The benchmark constructs artificial data in
which we can independently control relationships between the entities. It also provides a way to change relationships smoothly in time if needed. In this way, we
emulate correlations present in the real data, which are the main source of statistical artifacts. We use the benchmark to test the method and to find
applicable regimes in which extracted relationships are reliable, such as the level
of significance, the number of documents etc. A future envisioned application of the benchmark is to
compare different methods of relationship extraction from dynamic bipartite
networks and to use it as a testing ground for investigation of more sophisticated
methods. Evaluation of link importance in data and especially in temporal networks is in its infancy and we concluded that it would be reasonable to provide some way to create synthetic data which could be used to compare the performance of different algorithms as they are presented to the community.

We propose a simple benchmark model which creates a series of artificial document nodes with a time stamp and a list of four entities attached to it. The simplest case to calculate would be if we assign only two entities per document, but such a benchmark would lack correlations. Namely, if documents contain more than two entities, entities with several strong relationships will have in general more co-occurrences even with entities they have no relationship with. Documents with more than two entities introduce correlations as an immanent part of real data, and we also have to incorporate them in our benchmark. The choice of four entities per document is a compromise between simplicity and correlations in the data. Further in the text we provide formulae for the general model with $p$ different entities per document.

\subsection*{Benchmark model with 4 entities} 
 
In the benchmark we independently control each pair of entities by assigning them \emph{importance} $w_{ij}$ representing hidden relationships. Importance is in general a positive real number. The probability for a document to contain entities $i$, $j$, $k$ and $l$ we write as a function of all six importances (pairwise). 
\begin{equation}
P(\{i,j,k,l\}) = f(w_{ij},...,w_{kl})
\end{equation}
We use ``$\{$'' and ``$\}$'' to stress that the ordering of elements is not important and to distinguish these probabilities from the later use of probabilities in which the ordering is important.
In principle it is possible to calculate probabilities for all possible combinations of four entities but it seems to be rather expensive since one would have to calculate $\binom{N}{4}$ numbers, where $N$ is the number of entities. To be more efficient we develop an algorithm which picks entities one by one and still keep above probability. Such procedure requires only $4N$ calculations.\\

The simplest choice for the document probability is a sum of all six importances and we will adopt it as a reasonable choice. 
\begin{align}
P(\{i,j,k,l\}) = \frac{w_{ij}+w_{ik}+w_{il}+w_{jk}+w_{jl}+w_{kl}}{N_{\{w\}}}
\end{align}
where $N_{\{w\}}$ is a normalization factor. It should be mentioned that other choices like product rules or some other rule could also be devised and they could possibly be more realistic. However, for a more realistic function $f$ we would need to have a model of how the choices of entities are made by writers and we are not aware of any such models.
Since importances are not otherwise defined we can use this formula as their definition when interpreting the results.

When simulating benchmarks we want to avoid choosing randomly among sets of four entities - $\binom{N}{4}$ combinations. We reduce the problem to choosing entities one by one - four times $N$ combinations. For this we need probabilities:
\begin{enumerate}
\item $P(i,*,*,*)$ - probability to choose $i$ as first
\item $P(j|i,*,*,*)$ - probability to choose $j$ as second, given $i$ is first
\item $P(k|i,j,*,*)$ - probability to choose $k$ as third, given the first two
\item $P(l|i,j,k,*)$ - probability to choose $l$ as fourth, given the first three
\item $P(i,j,k,l)$ - probability to choose $i$,$j$,$k$ and $l$ in that order
\end{enumerate}

When constructing a document, probability for a \emph{first} entity to be $i$ is simply:
\begin{equation}
P(i) = \sum\limits_{\{j, k, l\}} \frac{P(\{i, j, k, l\})}{4}
\end{equation}
Division by 4 is the consequence of unordered character of the distribution $P(\{\cdots\})$.
To calculate probability for a second entity to be $j$ given we already have $i$ we first need to calculate the probability of $i$ and $j$ to be in the same document:
\begin{equation}
P(\{i,j\}) = \sum\limits_{\{k,l\}}P(\{i, j, k ,l\}).
\end{equation}
The probability that $i$ was first picked and $j$ second is $P(i,j)=P(\{i,j\})/12$
and using Bayes formula for conditional probability we have:
\begin{equation}
P(j|i) = \frac{P(i,j)}{P(i)}
\end{equation}
This procedure can now recursively done until we select all four entities.

Later we calculated the probabilities for a general number of entities in the document but we report the exact probabilities for the case with $4$ entities per document. 
\begin{align}
P(\{i,j,k\}) &= \frac{1}{N_{\{w\}}}((N-5)(w_{ij}+w_{ik}+w_{jk}) + s_i + s_j + s_k)\\
P(\{i,j\}) &= \frac{1}{2 N_{\{w\}}}((N-5)(N-4)w_{ij} + 2(N-4)(s_i+s_j)+\mathcal{S})\\
P(\{i\}) &= \frac{(N-3)}{2N_{\{w\}}}(s_i(N-4) + \mathcal{S})
\end{align}
where $s_i = \sum_k w_{ik}$ is a strength of the entity $i$ and $\mathcal{S} = \sum_i s_i$.

Normalization $N_{\{w\}}$ can be calculated to be:
\begin{align}
N_{\{w\}} &= \frac{1}{4}(N-2)(N-3)\mathcal{S} 
\end{align}
Therefore, we can calculate probabilities needed to pick entities one by one:
\begin{align}
P(i) &= \frac{1}{2(N-2)\mathcal{S}}((N-4)s_i + \mathcal{S})\\
P(j|i) &= \frac{1}{3(N-3)}\frac{(N-5)(N-4)w_{ij} + 2(N-4)(s_i+s_j)+\mathcal{S}}{(N-4)s_i + \mathcal{S}}\\
P(k|i,j) &= \frac{(N-5)(w_{ij}+w_{ik}+w_{jk}) + s_i + s_j + s_k}{(N-5)(N-4)w_{ij} + 2(N-4)(s_i+s_j)+\mathcal{S}}\\
	P(l|i,j,k) &= \frac{w_{ij}+w_{ik}+w_{il}+w_{jk}+w_{jl}+w_{kl}}{(N-5)(w_{ij}+w_{ik}+w_{jk}) + s_i + s_j + s_k}
\end{align}

For constant importances we do not need time stamps on documents and create a bipartite network with $N_D$ documents on which various methods of data extraction can be tested.
If, on the other hand, we want to test the time resolution of such methods, we can give importances time dependence; then time stamps are created as a realization of some random process in time.

\subsection*{General benchmark model}

Although in this paper we are using the variant of the model in which there are only $4$ entities in each document, it is easy to generalize the results to documents with $p$ different entities in each document.
In that case
\beq\label{GenPw}
P(\{i_1,\ldots,i_p\})=\frac{1}{N_{\{w\}}}\sum_{i_{\alpha},i_{\beta}\in\Omega_p}w_{i_{\alpha}i_{\beta}},
\eeq
is the probability that the document with $p$ entities will have all entities from the set $\{i_1, ..., i_p\}\equiv\Omega_p$. The probability that a subset of $r$ entities $\Omega_r\equiv\{i_1,...,i_r\}$, $\Omega_r\in\Omega_p$ will be found in the randomly chosen document is:
\beq\label{Pr1}
P(\{i_1,\ldots,i_r\})=\sum_{i_{r+1},...,i_{p}\in\bar{\Omega}_N}P(\{i_1,\ldots,i_r,i_{r+1},\ldots,i_p\}),
\eeq
and the set $\bar{\Omega}_N$ is a set of all the possible entities that are not found in the set ${i_1,\ldots,i_r}$. We will define one other set $\Omega_s=\Omega_p\backslash\Omega_r$ which is a set of entities that are contained in the document but are not in the set of entities whose probability of occurrence we calculate. Note that sets $\Omega_r$ and $\bar{\Omega_N}$ have fixed values of indices while set $\Omega{s}$ has variable indices that are elements of $\bar{\Omega_N}$ set.

Using equation (\ref{GenPw}) and equation (\ref{Pr1}) we write
\beq\label{Pr2}
P(\{i_1,\ldots,i_r\})=\frac{1}{N_{\{w\}}}\sum_{\Omega_s\subset \bar{\Omega}_N}\sum_{\substack{i_{\alpha},i_{\beta}\\ \in\Omega_p}}w_{i_{\alpha}i_{\beta}}.
\eeq
Note that here the first summation runs over all possible subsets $\Omega_s$ of the set $\bar{\Omega}_N$ and the second runs over all elements of chosen set $\Omega_p=\Omega_r\cup\Omega_s$. We can break the equation (\ref{Pr2}) into a sum of three distinct parts:
\beq\label{Pr3}
P(i_1,\ldots,i_r)=\frac{1}{N_{\{w\}}}\left(W_r+W_{rs}+W_s\right),
\eeq
where 
\beq
W_r\sum_{\Omega_s\subset \bar{\Omega}_N}\sum_{\substack{i_{\alpha},i_{\beta}\\ \in\Omega_r}}w_{i_{\alpha}i_{\beta}},\eeq 
is a contribution of links connecting the entities of the set $\Omega_r$; 
\beq 
W_{rs}=\sum_{\Omega_s\subset \bar{\Omega}_N}\sum_{\substack{i_{\alpha}\in\Omega_r,\\i_{\beta}\in\Omega_s}}w_{i_\alpha i_\beta}
\eeq 
is a contribution of links connecting the entities in a set $\Omega_r$ with elements in all possible sets $\Omega_s$ and 
\beq
W_s=\sum_{\Omega_s\subset \bar{\Omega}_N}\sum_{\substack{i_{\alpha},i_{\beta}\\ \in\Omega_s}}w_{i_{\alpha}i_{\beta}}
\eeq
is a contribution of links that are connecting entities all possible sets $\Omega_s$.

These contributions can be calculated as follows:
\begin{eqnarray}
 W_r&=&\binom{\bar{N}}{s}T_r,\nonumber\\
 W_{rs}&=&\binom{\bar{N}-1}{s-1}\left(\sum_{i_\alpha\in\Omega_r}s_{i_\alpha}-2T_r\right),\nonumber\\
 W_{s}&=&\binom{\bar{N}-2}{s-2}\left(\frac{\mathcal{S}}{2}-\sum_{i_\alpha\in\Omega_r}s_{i_\alpha}+T_r\right),
\end{eqnarray}
where $\bar{N}=N-r$ and $T_r=\sum_{i_{\alpha},i_{\beta}\in\Omega_r}w_{i_{\alpha}i_{\beta}}$. Now we can write equation (\ref{Pr3}) as:
\begin{eqnarray}
 N_{\{w\}} P(\{i_1,\ldots,i_r\})&=&\left(\binom{\bN}{s}+\binom{\bN-2}{s-2}-2\binom{\bN-1}{s-1}\right)T_r\nonumber\\
 &&+\binom{\bN-2}{s-2}\frac{\mathcal{S}}{2}\nonumber\\
 &&+\left(\binom{\bN-1}{s-1}-\binom{\bN-2}{s-2}\right)S_r,
\end{eqnarray}
where $S_r=\sum_{i_\alpha\in\Omega_r}s_{i_\alpha}$. This equation can be written in a more condensed form as:
\beq
  P(\{i_1,\ldots,i_r\})=\frac{\binom{\bN-2}{s-2}}{N_{\{w\}}}\left(\frac{(\bN-s)(\bN-s-1)}{s(s-1)}T_r+\frac{\bN-s}{s-1}S_r+\frac{\mathcal{S}}{2}\right)
\eeq
Once the probabilities are computed it is easy to compute ordered probabilities needed for the computer simulation. Probability $P(i_1,...,i_r)$ that first the $i_1$ was chosen then $i_2$ and all the way to $i_r$ is just:
\beq
P(i_1,...,i_r)=\frac{P(\{i_1,...,i_r\})(p-r)!}{p!}.
\eeq
Note that this means that we can also work with ordered probabilities $P(i_1,...,i_r)$ for which
\beq
P(i_1,...,i_p)=\frac{1}{N_{w}}\sum_{i_{\alpha},i_{\beta}\in\Omega_p}w_{i_{\alpha}i_{\beta}},
\eeq
with normalization factor $N_{w}=N_{\{w\}}\cdot p!$.
A more detailed version of general calculations is available in the appendix.

\section*{Results}

In this section we present the evaluation of the proposed algorithm for estimating the significance of entity co-occurrences. First the results obtained on synthetic data are presented, followed by a comparison of temporal networks constructed from country co-occurrences in financial news and from correlations between the corresponding CDS time series.

\subsection*{Testing the method with syntethic data}
We tested our method on artificial networks produced with the benchmark model. We use artificial networks to estimate the statistics needed to reliably construct financial
interdependence networks from the available data. Such networks can then be used as
proxies for real financial networks.

We test two types of artificial networks created by the benchmark model, both with $N = 100$ nodes representing entities and $L=4950$ potential links in the projection network. The number of simulated documents $N$ is between $10^2$ and $10^4$. 

\begin{figure}[!ht]
\begin{center}
\includegraphics[width=4in]{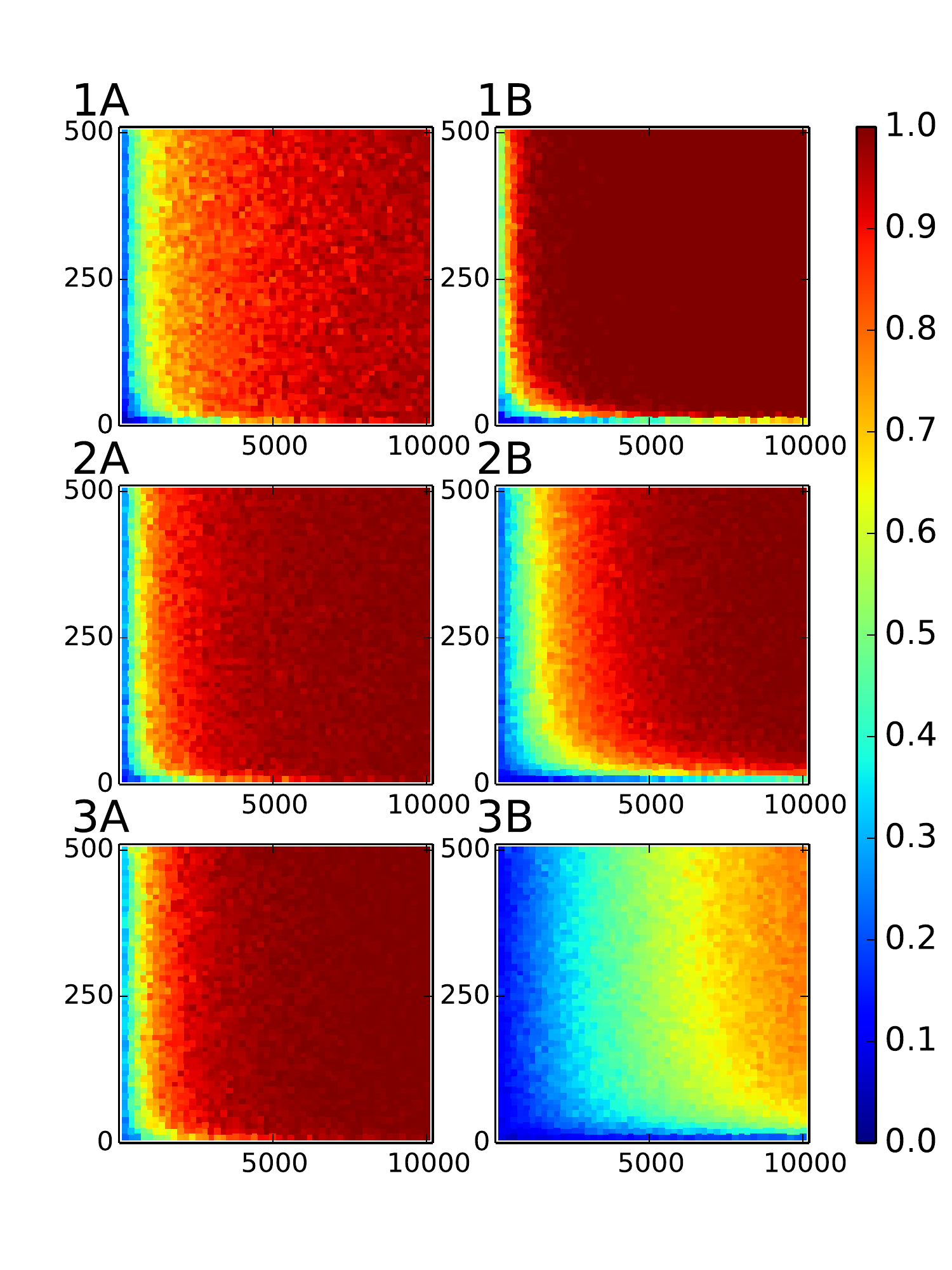}
\end{center}
\caption{{\bf PPV and sensitivity for the case of two types of importances}. The x-axis gives the number of documents and the y axis is importance amplitude. In the column A colors show positive prediction value and in the column B colors represent sensitivity. The number of important links is $N_{\text{significant}} = 100$ for the first row, $N_{\text{significant}} =  250$ for the second row and $N_{\text{significant}} =  500$ for the third row.}
\label{importances}
\end{figure}

The first aforementioned type is constructed with $k$  importances $w_{ij}$ randomly drawn from the set of two different values $\{w_0,w_I\}$, satisfying the constraint that the number of importances $w_{ij}$ with assigned value $w_I$ is exactly $N_{significant}$. Thus $L-N_{significant}$ relations have assigned weights $w_0 = 1$ and  $N_{significant}$ important relations have assigned weights $w_I=aw_0$, where $a$ is a parameter we call importance amplitude $a > 1$. Using the method to determine the significant links with the threshold $Z_0 = 2.0, 3.0, 5.0$ on the realizations of artificial networks, we can calculate a fraction of significant links that are also important, i.e. $w_{ij}=w_i$ (positive predictive value - PPV) and a fraction of important links found to be significant (sensitivity). See figure \ref{importances}. The number of significant links in the figure is $N_{\text{significant}} = 100, 250, 500$, with a variable number of documents $N$, and importance amplitude $a$.

Figure \ref{importances} show that 3000-5000 documents are good enough to provide reliable significance for extracted co-occurrences as long as importance is high enough. If the difference between an important link and an average link is not large enough there is no number of documents that we tested that will result with in high PPV or sensitivity.

\begin{figure}[!ht]
\begin{center}
\includegraphics[width=4in]{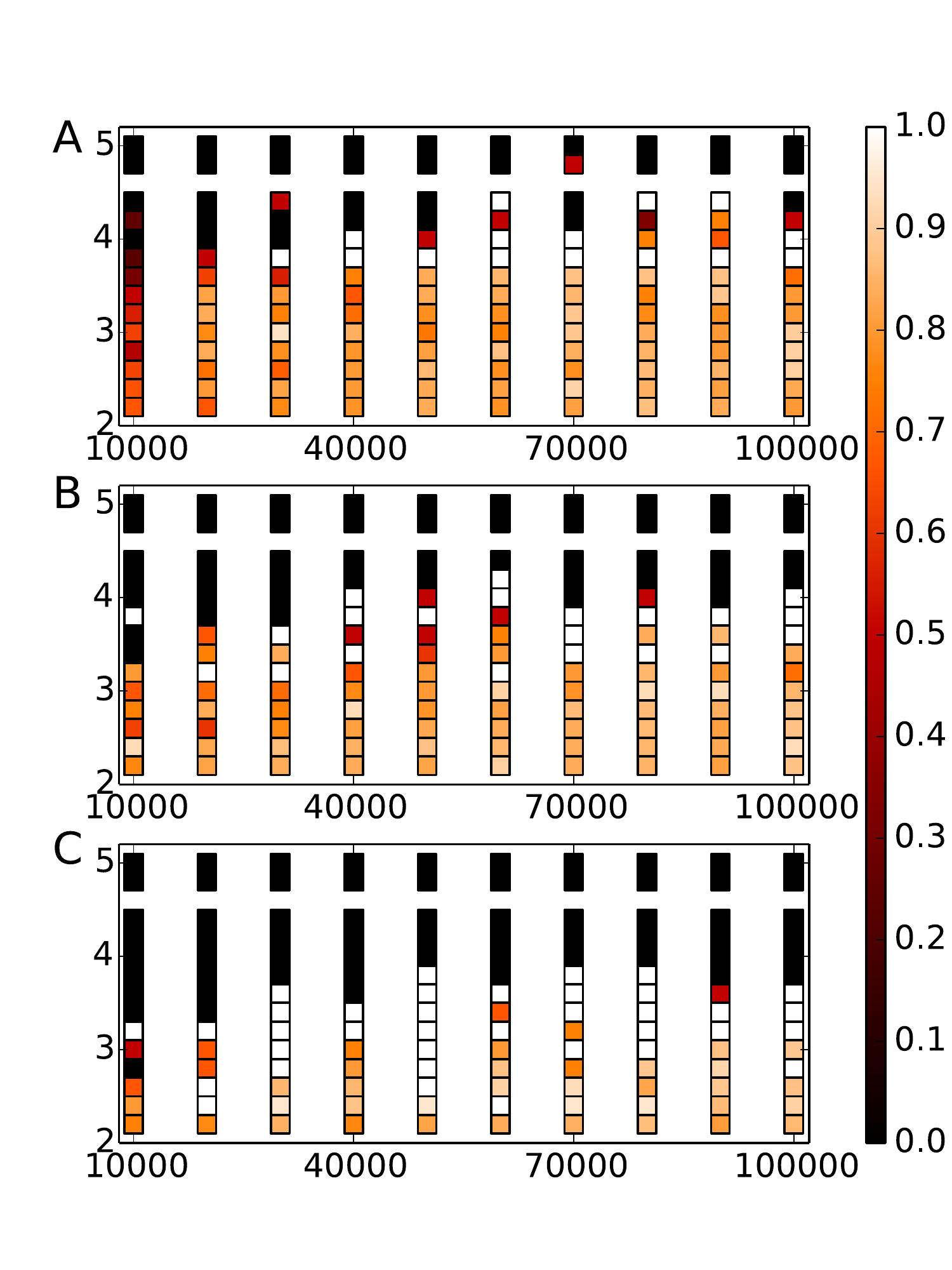}
\end{center}
\caption{{\bf PPV for the case of power law distributed importances.} The x-axis gives the number of documents and the y-axis is exponent of the power law for the second case. Colors show positive prediction value. The threshold is $Z_0 = 2.0$ for the panel A, $Z_0 = 3.0$ for the panel B and $Z_0 = 5.0$ for the panel C. }
\label{importances2}
\end{figure}

In the second type of artificial networks, importance is a power law distributed with exponent $\gamma$. We determine significant links with some threshold $Z_0$. The number of such links is $N_s$. To measure the number of correctly selected links, we compare the list of significant links to the list of $N_s$ links with the highest importance. Links occurring in both lists are true positives and their fraction with respect to the $N_s$ is a measure of the method's performance. See figure \ref{importances2} for values of $Z_0 = 2, 3, 5$.

It is clear that the method works well for exponent $\gamma\lesssim 4$. The reason is that for higher exponents, important links do not have a large spread i.e., typical importance in the data set is close to the highest importance in the data set, and we can see again that the number of documents cannot improve statistics much. The correlations between the co-occurrences imposed by the number of entities per document in the data set are masking the real importance of the links.

\subsection*{Comparison of co-occurrence networks with CDS networks}
The final goal of co-occurrence networks is to provide some information on the relationships between entities of interest. We hope that the co-occurrence networks of financial entities can provide information about shared risk of the entities of interest.
The idea is to compare data indirectly related to sovereign debt.
A temporal network was extracted from co-occurrences
of countries in financial news, and another was constructed from the correlations between CDS time series of
the same countries.

In the financial literature\cite{pan2008default, aizenman2012risk},
CDS are often considered a good proxy for the risk of default of a financial institution issuing bonds. The structure of that financial instrument is "triadic" in the sense that a CDS is a special insurance policy that a financial institution sells (seller) or buys (buyer) to hedge against the risk that a third party (reference entity) will experience a default within some fixed period (the maturity of the CDS), and the financial investments of the buyer will be lost. In the formation of the price the triadic model (that accounts for the risk of the buyer, the seller and the reference entity) will produce, at the end of computation, a single value (the CDS price) for each financial institution.
This number is considered an estimation, from a market perspective, of the perceived default risk but, as the CDS are financial products, the dynamics of the prices can also follow other market trends that are not immediately bound to systemic risk. We cautiously suggest that the CDS time series are a possible proxy of the systemic risk of a country as this concept can involve many other components in addition to financial ones. The networks are reconstructed using the correlation of pairs of time series, one per country. We recall here that a high correlation between country A and country B, during a given period, does not necessarily imply an high risk as it is also important to consider the common level of the prices, i.e. during normal business we can have high correlation but low risk while, conversely, during bad business we can see small correlation and high risk. We conclude that the correlation networks account for similar patterns in CDS prices across different countries while the mapping with the 
systemic risk needs then to be clearly stated from the price levels and possibly from other financial indicators. 

The entity co-occurrence network was constructed from textual data in the form of financial news and blogs from November 1, 2011 until December 31, 2013.  In this period, the acquisition pipeline collected about 18 million documents. They were filtered for strictly financial news, and each document had to contain at least two different entities, each occurring at least twice. This filtering resulted in more than 1.3 million documents to be analyzed. We chose the observed entities to be 50 selected \emph{countries} and the corresponding economic indicators to be the countries' \emph{CDS} time series.
We were hoping to see that countries having a higher correlation in CDS prices tend to be cited together in the news: higher correlation is reflected in the media as more co-occurrences.


Links in the country co-occurrence network were created according to the method for extraction of important co-occurrences, as presented in the Significance Algorithm section. The links in the CDS network were created using the Pearson's correlation coefficient ($c$) among two CDS time series. The temporal networks were constructed using a rolling window of three months that was shifted for one month over a period of two years. This time window was chosen using the benchmark model so that the average number of documents is large enough to reliably detect hidden relationships.

Our comparison of networks constructed from significant co-occurrences and CDS correlations was twofold. First, we examined the overlap of the most important links in the networks, and second, we compared the structure of both networks by investigating the similarity between their most central nodes.

\begin{figure}[!ht]
\begin{center}
\includegraphics[width=4in]{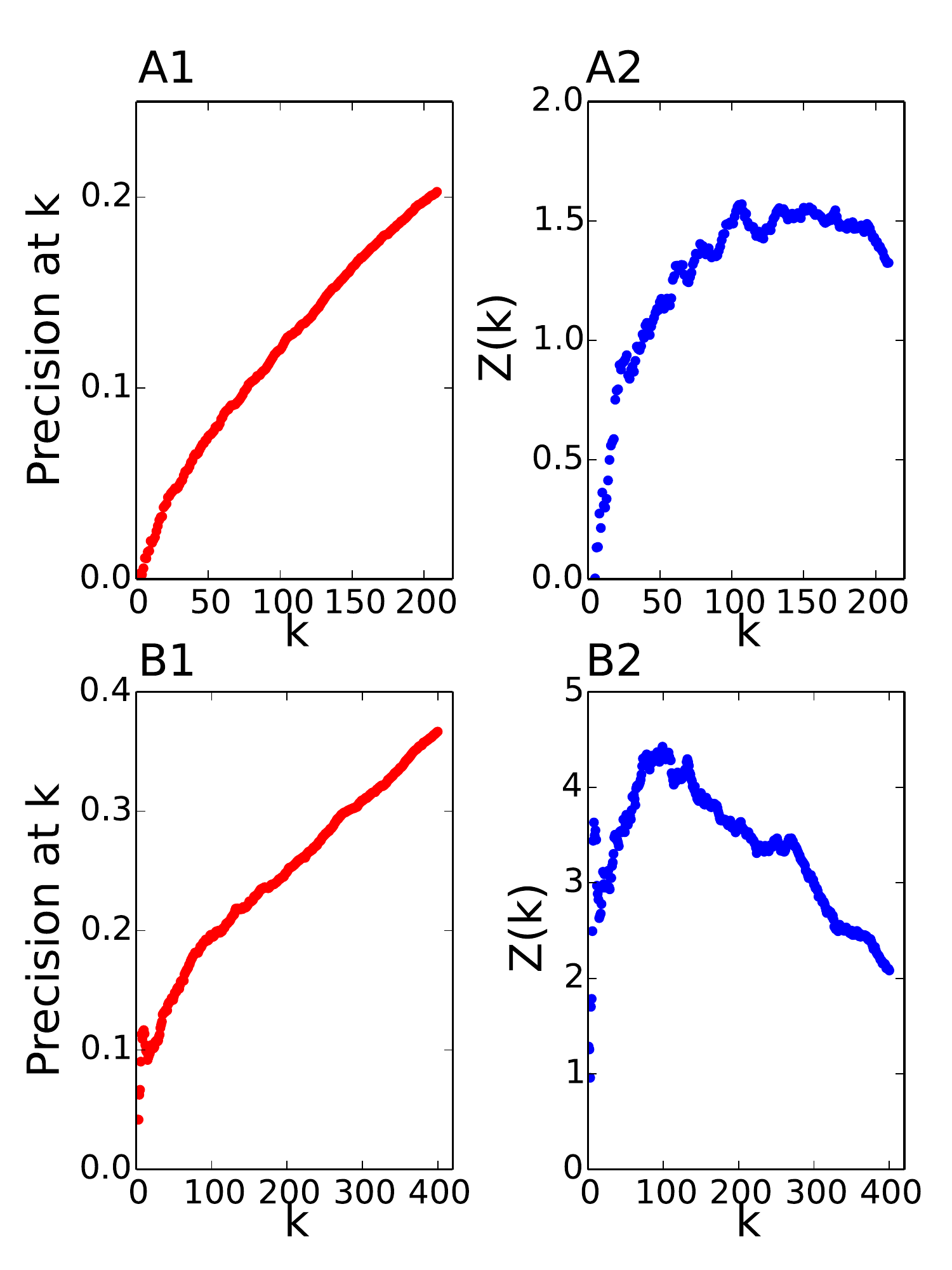}
\end{center}
\caption{{\bf Precision at k for the matching between co-occurrence networks and the CDS networks} On the left panels is precision at level k and on the right associated z score produced from comparison of the data with 10.000 randomized versions of data. Top is the matching between two networks with 1 week integration period and bottom is matching between 2 networks with 3 months integration period. }
\label{PrecisionAtK}
\end{figure}

\begin{figure}[!ht]
\begin{center}
\includegraphics[width=4in]{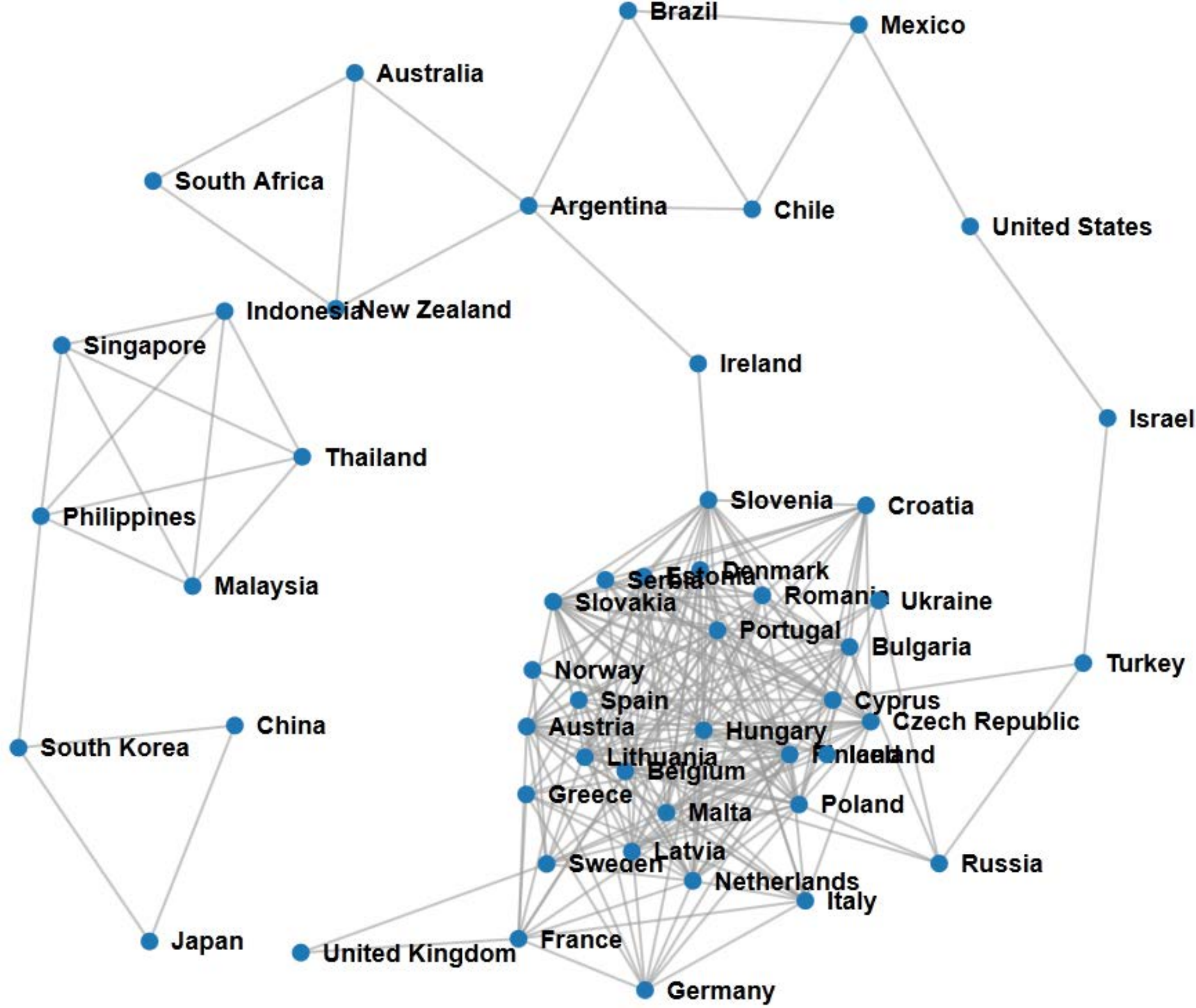} 
\end{center}
\caption{{ \bf Co-occurrence network created with $Z_0=15$ for October 2012.}}
\label{Network1}
\end{figure}

\begin{figure}[!ht]
\begin{center}
\includegraphics[width=4in]{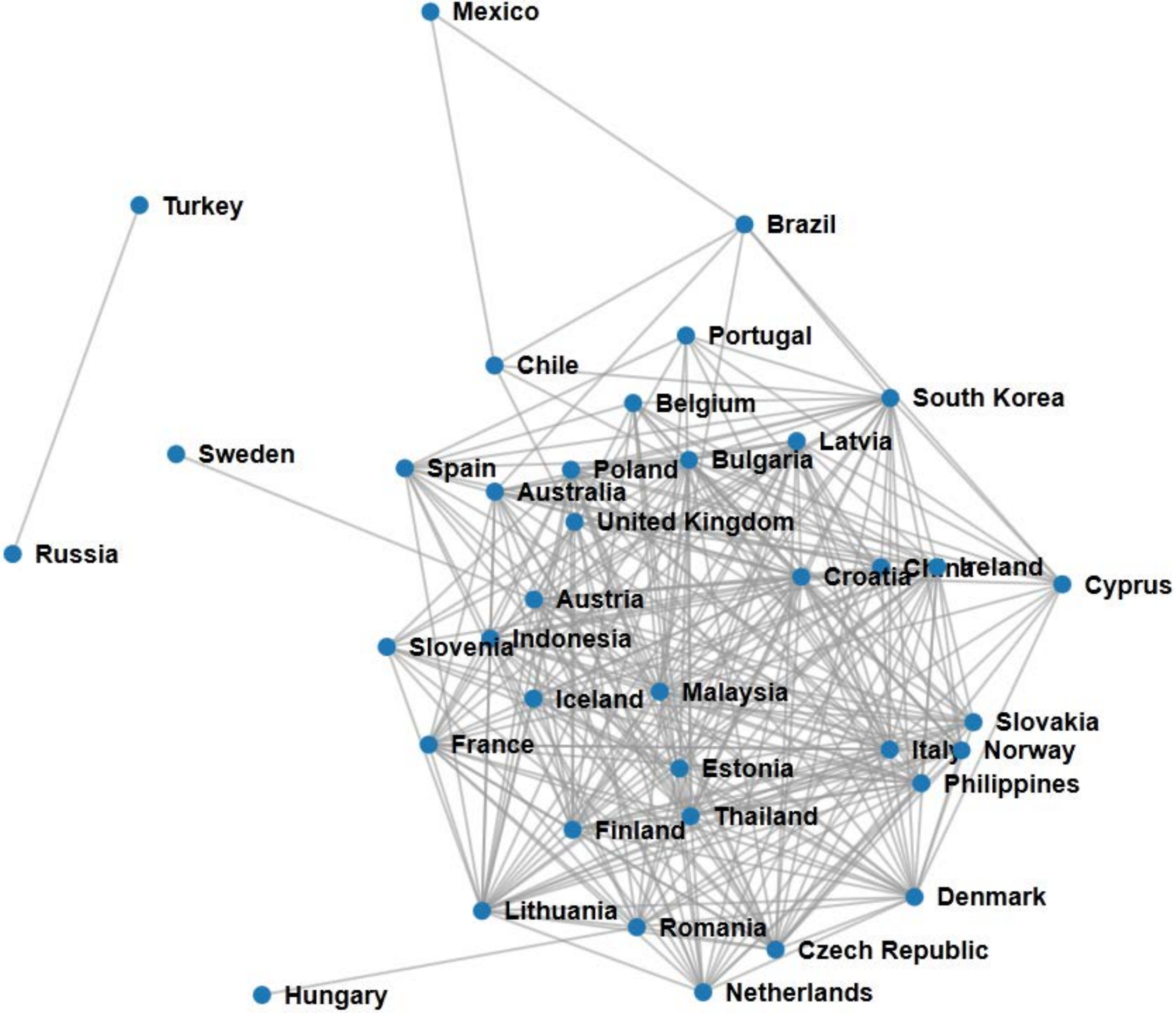} 
\end{center}
\caption{{ \bf CDS network created with Pearson's correlation $>0.9$ for October 2012.}}
\label{Network2}
\end{figure}

To compare the networks in terms of their most important links we have used the \emph{precision at $k$} method \cite{raghavan1989critical}, commonly used as a metric in recommender systems. Precision at $k$ is defined as follows. First, the links in both networks are ordered by their importance. In the case of co-occurrences we use the significance computed with our method and for the CDS networks we use the correlations as measures of importance. Then we count how many links are present in both ordered lists in the first $k$ entries. Finally, the precision at $k$ is defined as the fraction of the matched links  
\beq
P(k)=\frac{Number\;of\;matched\;links}{k}.  
\eeq
We present the results of matching links for networks constructed in two different sliding window settings: one week window sliding from week to week, and a three month window sliding by one month. The results are summarized in Figure~\ref{PrecisionAtK}. It is clear that in one week we are not able to collect statistics reliable enough to match the co-occurrence networks with the CDS networks, or that the relationship at such fine temporal scales does not exist. On the other hand, for 3-month integrated data we see that there is a significant match between co-occurrence networks and CDS networks. To evaluate the significance of the match we have used the $Z$-score value. To obtain the expected number of matchings and the standard deviation we have used 10.000 random permutations of ordered lists. The distribution of random matchings is Poisson like, which allows the use of a $Z$-score for significance testing. As can be seen in Figure~\ref{PrecisionAtK}, the matching for a one week window is really very 
small and only modestly better then completely random matching. On the other hand, the
matching 
for the 3 month window is significant with a signal easily surpassing $4\sigma$ between $k\cong 50$ and $150$. One other possible interpretation for this result is that in our case study we have used country CDS data which are less volatile than the companies. We did not use companies in order to get good enough statistics, as evaluated with the benchmark model, for the comparison with CDS data. Examples of the co-occurrence network constructed with our method and the CDS network for the same period are presented in Figures~\ref{Network1} and ~\ref{Network2}, respectively.

\begin{figure}[!ht]
\begin{center}
\includegraphics[width=0.9\linewidth]{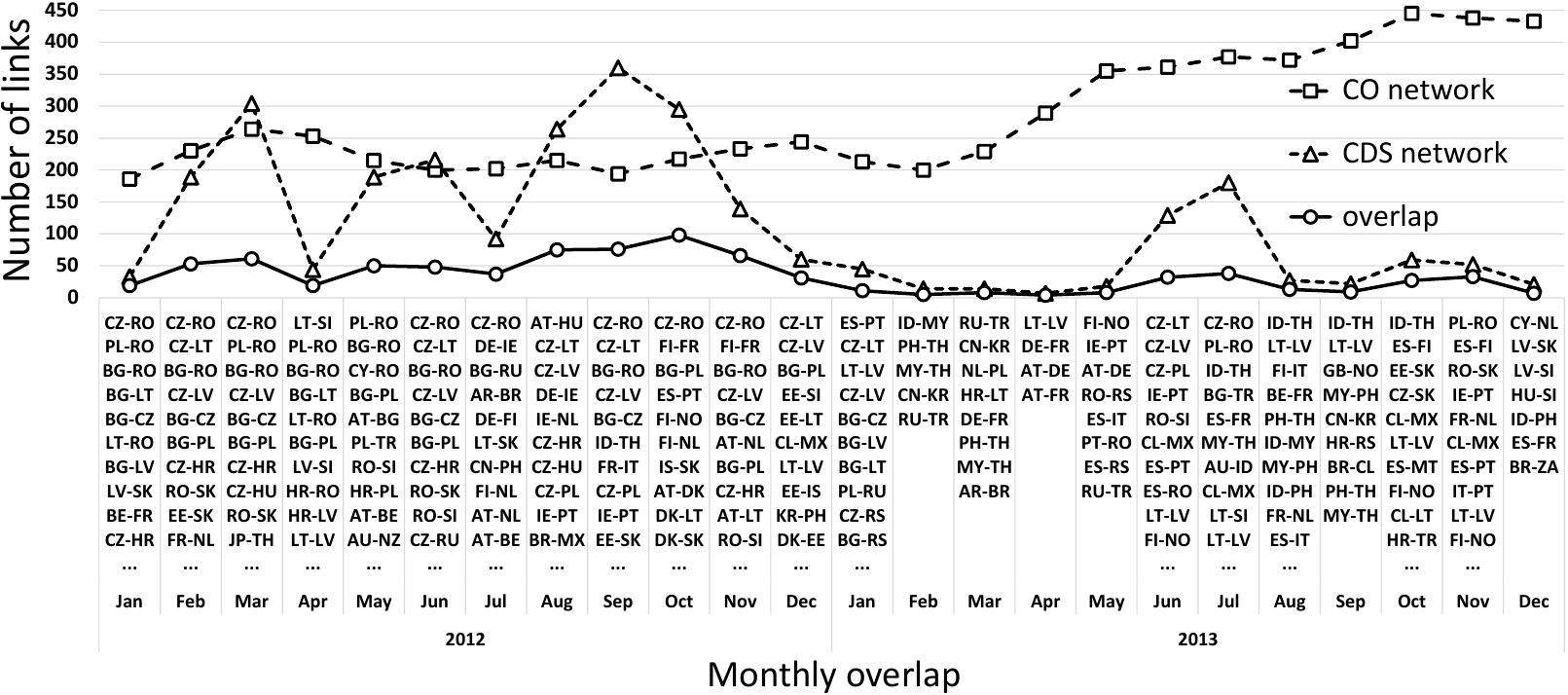}
\end{center}
\vspace*{-0.5cm}
\caption{{\bf Monthly overlapping links of the co-occurrence and CDS networks, for $Z > 15$ and $c > 0.9$. Ten most significant links in the monthly overlaps are listed.}}
\label{LinkOverlaps}
%
\begin{center}
\includegraphics[width=0.9\linewidth]{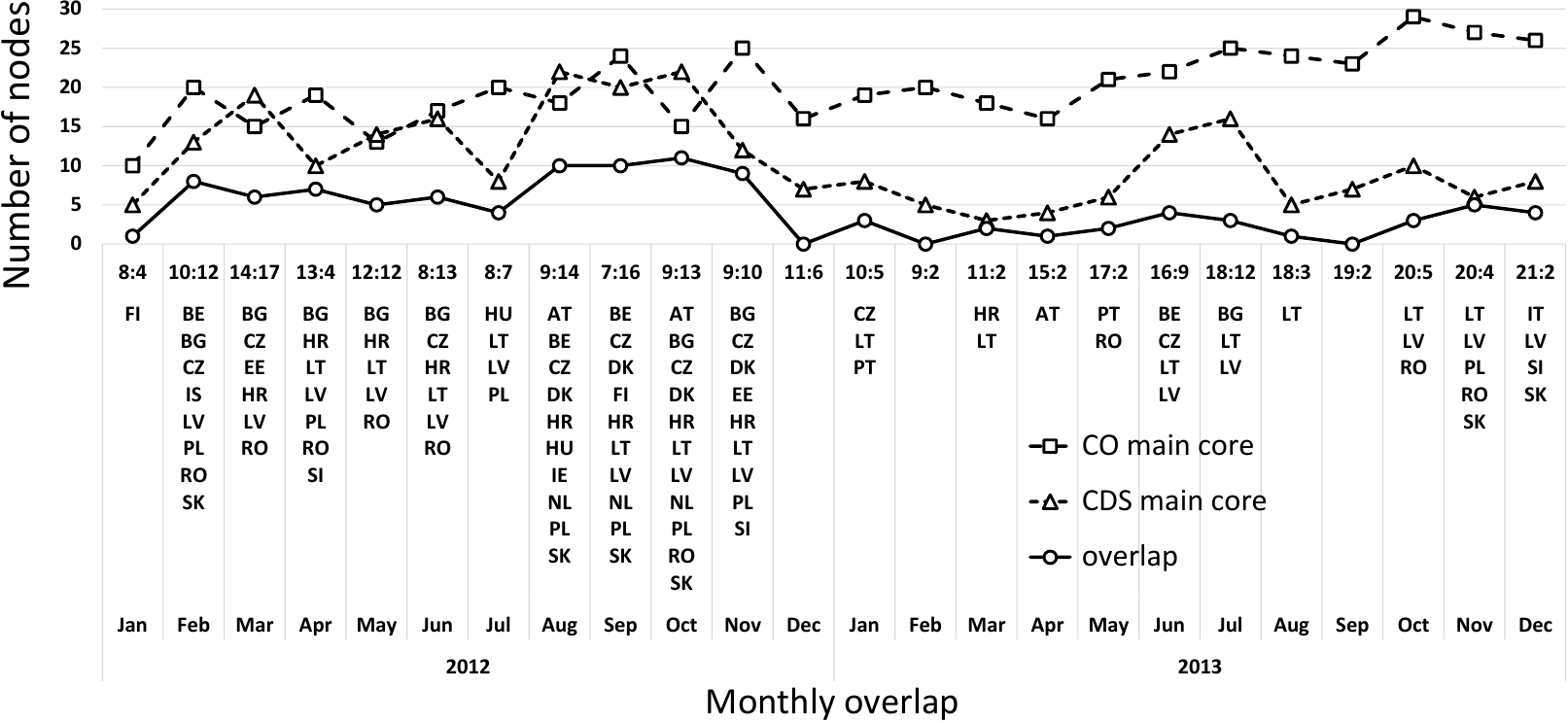}
\end{center}
\vspace*{-0.5cm}
\caption{{ \bf Monthly overlapping nodes of the co-occurrence and CDS networks’ main cores.} The respective coreness is indicated in the form $k_{CO}:k_{CSD}$.}
\label{kCoreNodeOverlaps}
%
\begin{center}
\includegraphics[width=0.9\linewidth]{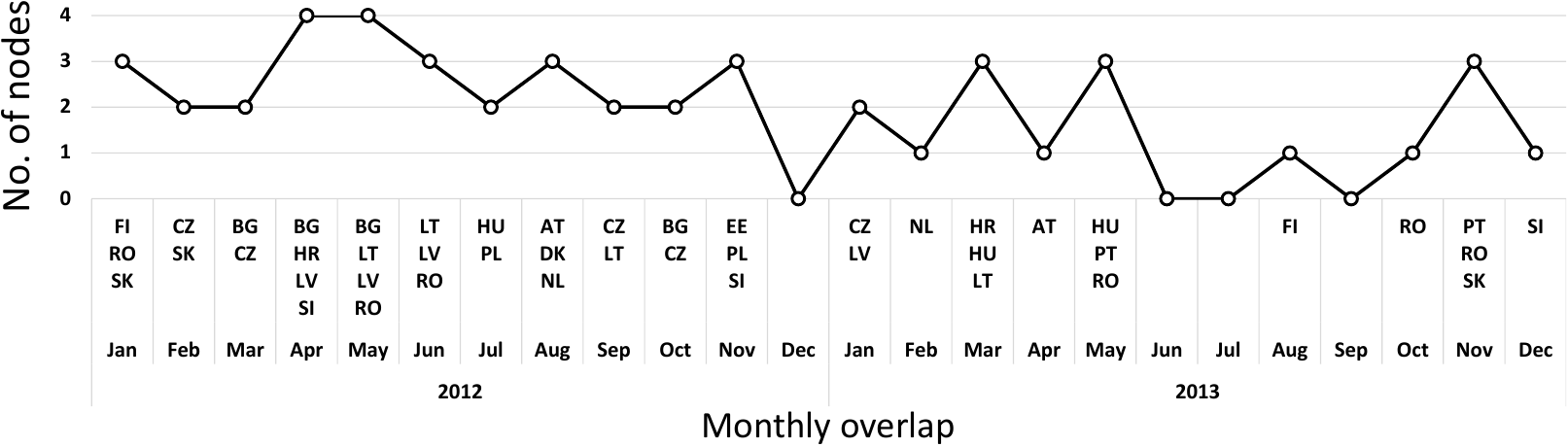}
\end{center}
\vspace*{-0.5cm}
\caption{{ \bf Monthly overlaps of the ten most central nodes from the co-occurrence and CDS networks.} Note that most of the overlapping noted belong also to the overlaps of the main network cores as shown in Figure~\ref{kCoreNodeOverlaps}.}
\label{centralNodeOverlaps}
\end{figure}

The networks shown in Figures~\ref{Network1} and~\ref{Network2} are constructed from the links between the most significantly co-occurring countries, $Z > 15$, and the highest correlating CDS time series, $c > 0.9$. In Figure~\ref{LinkOverlaps} we show the monthly overlaps of links in the co-occurrence and CDS networks over the period of two years. Among the most important links that the two networks share we can observe many pairs of country names that are known to be connected economically, as well as some geographically (which in some cases implies similar economic indicators).

Another way to compare the structure of two networks is by looking at the most important nodes that they have in common. We performed a $k$-core decomposition~\cite{seidman1983kcore} of the networks and compared the overlap between their main cores, i.e. the $k_{\max}$-cores of the respective networks. The monthly overlaps and the lists of nodes in the overlaps are presented in Figure~\ref{kCoreNodeOverlaps}. The results show a moderate level of overlap between the main core nodes, but provide insight into commonly and repeatedly appearing nodes in the overlap. Additionally, we examined the most central nodes in the networks, as denoted by the eigenvector centrality measure~\cite{bonacich1972evcentrality}. The monthly overlaps of ten most central nodes in the co-occurrence and CDS networks are presented in Figure~\ref{centralNodeOverlaps}. The overlap between the most central nodes is also in this case moderate, but shows that similar nodes are most important as in the overlap analysis with the $k$-core 
decomposition. Note that most of the overlapping nodes in Figure~\ref{centralNodeOverlaps} belong also to the overlaps of the main network cores as shown in Figure~\ref{kCoreNodeOverlaps}.

\section*{Conclusion}

The method presented in the paper is simple and fast and therefore well suited for implementation of fast significance detection in huge streams of data. We have presented and implemented a pipeline for real-time
acquisition and analysis of a stream of financial news. The extraction of significant co-occurrences was tested on historical data, but can
be added to the real-time processing pipeline.
More sophisticated methods could outperform this method in realistic settings given sufficient time. 

We have also presented a method for the creation of bipartite networks based on a hidden variable model with given importances. We have tested our model on synthetic data which we produced with the benchmark model and used it to find a time window which will, on average, have enough documents for a reasonable reconstruction of entity relationships. Furthermore, every new method can be tested with this benchmark to assess its validity and performance. It is important to stress that the benchmark model can easily be extended to include time-changing importances in order to test the statistics needed to capture the change in the value of importances and so on. 
Furthermore, we are preparing a manuscript in which we will use a generalization of this method for creation of ``canonical'' models of bipartite networks. Since all the relationships are linear it is possible to invert the matrix that relates importances $w_{ij}$'s to co-occurrence distribution $P(\{i,j\})$ and to extract from the data importances for which expected distribution of co-occurrences is exactly the one found in the data.

We have also shown that in the cases of large enough datasets we can relate co-occurrence networks with networks of mutual financial risk, such as CDS networks. Our method was thoroughly investigated by means of the benchmark model for the reliable number of events that could discover relationships between news data and CDS market. Our results show that the relationship between two data sets is significant but very weak. This may be attributed to several causes. \emph{(i)} One possibility is that there is really no strong relationship between the news and the market. In this case further investigation of this relationship should yield similar results. \emph{(ii)} We have used only news in the English language. It is possible that English news is biased in such a way that only a small portion of the market is well presented. In this case future research should include news samples in many different languages to show a stronger relationship between news and the CDS market. \emph{(iii)} The relationship is 
hidden in longer time intervals and more pronounced in shorter time intervals. In that case a significantly larger sample of news should be used in order to provide for more reliable statistics in the shorter time intervals. Further work on the applicability of this method and causal relationships between CDSs and co-occurrences is proposed for the future. 

\section*{Appendix}

If each document contains $p$ entities, the probability of finding a set of entities $\{i_1,i_2,\dots,i_p \}$ is
\begin{equation}
\label{eq:defp}
P(\{i_1,i_2,\dots,i_p\})=\frac{1}{N_{\{w\}}} \sum_{1 \le \alpha < \beta \le p} w_{i_{\alpha} i_{\beta}} \, .
\end{equation}

The probability of finding a subset of entities $\{i_1,i_2,\dots,i_r \}$, where $r<p$, then reads
\begin{equation}
\label{eq:defr}
P(\{i_1,i_2,\dots,i_r\}) = \sum_{\{i_{r+1}, \dots,i_p\}} P(\{i_1,i_2,\dots,i_p\})= \frac{1}{N_{\{w\}}} \sum_{\{i_{r+1}, \dots,i_p\}}  \sum_{1 \le \alpha < \beta \le p} w_{i_{\alpha} i_{\beta}} \, .
\end{equation}
Here the sum over indices $\alpha$ and $\beta$ in (\ref{eq:defp}) can be decomposed into three sums in which none, one or both indices in the $w_{i_{\alpha} i_{\beta}}$ is summed over.
\begin{equation}
\label{eq:defrpairs}
P(\{i_1,i_2,\dots,i_r\}) =  \frac{1}{N_{\{w\}}} \sum_{\{i_{r+1}, \dots,i_p\}} \left[ \sum_{1 \le \alpha < \beta \le r <  p} w_{i_{\alpha} i_{\beta}} + 
\sum_{1 \le \alpha \le r < \beta \le p} w_{i_{\alpha} i_{\beta}} + \sum_{r< \alpha < \beta \le p} w_{i_{\alpha} i_{\beta}} \right] \, ,
\end{equation}
In particular these sums are calculated as follows.
The sum term where none of the indices in $w_{i_{\alpha} i_{\beta}}$ is summed over reads:
\begin{eqnarray}
\label{eq:zerosum}
& & \sum_{\{i_{r+1}, \dots,i_p\}} \sum_{1 \le \alpha < \beta \le r <  p} w_{i_{\alpha} i_{\beta}} \nonumber \\
&=& \sum_{1 \le \alpha < \beta \le r <  p} \frac{1}{(p-r)!} \sum_{i_{r+1}}
\sum_{i_{r+2} \not = i_{r+1}} \dots \sum_{i_p \not = \{i_{r+1},\dots,i_{p-1}\}} w_{i_{\alpha} i_{\beta}} \nonumber \\
& = & \sum_{1 \le \alpha < \beta \le r <  p} \frac{(N-r)!}{(p-r)! (N-p)!} w_{i_{\alpha} i_{\beta}} \nonumber \\
&=& \left( \begin{array}{c} N-r\\ p- r \end{array} \right) \sum_{1 \le \gamma < \delta \le r} w_{i_{\gamma} i_{\delta}} \, .
\end{eqnarray}

The sum term where one of the indices in $w_{i_{\alpha} i_{\beta}}$ is summed over reads:
\begin{eqnarray}
\label{eq:onesum}
& & \sum_{\{i_{r+1}, \dots,i_p\}} \sum_{1 \le \alpha \le r < \beta \le p} w_{i_{\alpha} i_{\beta}} \nonumber \\
&=& \sum_{1 \le \alpha \le r < \beta \le p} \frac{1}{(p-r)!} \sum_{i_{\beta}}
\sum_{i_{r+1} \not = i_{\beta}} \dots \sum_{i_{\beta-1} \not = \{ i_{r+1}, \dots i_{\beta-2}, i_{\beta}\}} \sum_{i_{\beta+1} \not = \{i_{r+1}, \dots, i_{\beta} \}}
\dots \sum_{i_p \not = \{ i_{r+1},\dots,i_{p-1}\}}  w_{i_{\alpha} i_{\beta}} \nonumber \\
&=& \sum_{1 \le \alpha \le r < \beta \le p} \sum_{i_{\beta}}  \frac{(N-r-1)!}{(p-r)! (N-p)!} w_{i_{\alpha} i_{\beta}} \nonumber \\
&=& \left( \begin{array}{c} N-r-1 \\ p- r-1 \end{array} \right) \left( \sum_{1 \le \gamma \le r} s_{i_{\gamma}} -2 \sum_{1 \le \gamma < \delta \le r} w_{i_{\gamma} i_{\delta}} \right) \, .
\end{eqnarray}

The sum term where both indices in $w_{i_{\alpha} i_{\beta}}$ are summed over reads:
\begin{eqnarray}
\label{eq:twosum}
& & \sum_{\{i_{r+1}, \dots,i_p\}} \sum_{r< \alpha < \beta \le p} w_{i_{\alpha} i_{\beta}}  \nonumber \\
&=& \sum_{r< \alpha < \beta \le p} \sum_{i_{\alpha}} \sum_{i_{\beta} \not = i_{\alpha}} \sum_{i_{r+1} \not = \{i_{\alpha},i_{\beta} \}}
\dots \sum_{i_{\alpha-1} \not = \{ i_{r+1}, \dots i_{\alpha-2},i_{\alpha}, i_{\beta}\}} \sum_{i_{\alpha+1} \not = \{i_{r+1}, \dots, i_{\alpha},i_{\beta} \}} \nonumber \\
&\dots& \sum_{i_{\beta-1} \not = \{ i_{r+1}, \dots i_{\beta-2}, i_{\beta}\}} \sum_{i_{\beta+1} \not = \{i_{r+1}, \dots, i_{\beta} \}}
\dots \sum_{i_p \not = \{ i_{r+1},\dots,i_{p-1}\}}  w_{i_{\alpha} i_{\beta}} \nonumber \\
&=&  \sum_{r< \alpha < \beta \le p} \sum_{i_{\alpha}} \sum_{i_{\beta} \not = i_{\alpha}}  \frac{(N-r-2)!}{(p-r)! (N-p)!} w_{i_{\alpha} i_{\beta}} \nonumber \\
&=& \frac{1}{2} \left( \begin{array}{c} N-r-2 \\ p- r-2 \end{array} \right) \left({\cal S} -2 \sum_{1 \le \gamma \le  r} s_{i_{\gamma}} + 2 \sum_{1 \le \gamma < \delta \le r} w_{i_{\gamma} i_{\delta}} \right) \, .
\end{eqnarray}
Here ${\cal S}=\sum_{\alpha=1}^p s_{i_{\alpha}}$. Further introducing
\begin{equation}
\label{eq:sr}
S_r=\sum_{1 \le \gamma \le r} s_{i_{\gamma}}
\end{equation}
and 
\begin{equation}
\label{eq:tr}
T_r=\sum_{1 \le \gamma < \delta \le r} w_{i_{\gamma} i_{\delta}}
\end{equation}
and combining (\ref{eq:zerosum}), (\ref{eq:onesum}) and (\ref{eq:twosum}), the expression for (\ref{eq:defr}) becomes
\begin{equation}
\label{eq:defr2}
P(\{i_1,i_2,\dots,i_r\}) = \frac{1}{N_{\{w\}}} \left( \begin{array}{c} N-r-2 \\ p- r-2 \end{array} \right) \left[ \frac{(N-p)(N-p-1)}{(p-r)(p-r-1)} T_r +\frac{N-p}{p-r-1} S_r + \frac{1}{2} {\cal S} \right] \, .
\end{equation}
The expressions given above are valid when $p > r+1$ because only in these cases both $T_r$ and $S_r$ terms appear. The formula (\ref{eq:defr2}) is also applicable to the case $r=1$ if we take $T_1=0$ and to the case $r=0$ if we take $T_0=0$ and $S_0=0$.

Finally, the case $r=p-1$ requires separate approach. In particular,
\begin{eqnarray}
\label{eq:rpminus1}
& & P(\{i_1,i_2,\dots,i_{p-1}   \}) = \sum_{i_p \not = \{i_1,i_2,\dots,i_{p-1}\}} \frac{1}{N_{\{w\}}} \sum_{1 \le \alpha < \beta \le p} w_{i_{\alpha} i_{\beta}} \nonumber \\
& = & \sum_{i_p \not = \{i_1,i_2,\dots,i_{p-1}\}} \frac{1}{N_{\{w\}}} \left( \sum_{1 \le \alpha < \beta \le p-1} w_{i_{\alpha} i_{\beta}}  + 
\sum_{1 \le \alpha \le p-1} w_{i_{\alpha} i_p} \right) \nonumber \\
&=& \frac{1}{N_{\{w\}}} \left[(N-p-1)\sum_{1 \le \alpha < \beta \le p-1} w_{i_{\alpha} i_{\beta}} +  \sum_{1 \le \alpha \le p-1} s_{i_{\alpha}} \right]
\nonumber \\
&=& \frac{1}{N_{\{w\}}} \left[ (N-p-1) T_{p-1} + S_{p-1} \right] \, .
\end{eqnarray}
\section*{Acknowledgements}

This work was supported in part by the European Commission under the
FP7 projects FOC (Forecasting financial crises, grant no. 255987) and
MULTIPLEX (Foundational Research on MULTIlevel comPLEX networks 
and systems, grant no. 317532), and
by the Slovenian Research Agency programme Knowledge Technologies
(grant no. P2-103). We thank Sebastian Schroff from Stuttgart Stock
Exchange for providing the CDS data for selected countries. MP, HS and
VZ have worked on the method and benchmark model. BS, PKN, MG and IM
have worked on the data acquisition pipeline and BS and MP have worked
on the CDS networks. All authors have contributed to the text of the
paper.

\bibliography{plosone_paper}

\pagebreak
\section*{Figure Legends}

\end{document}